\documentclass[10pt,conference]{IEEEtran}
\IEEEoverridecommandlockouts
\usepackage{cite}
\usepackage{amsmath,amssymb,amsfonts}
\usepackage{algorithmic}
\usepackage{graphicx}
\usepackage{textcomp}
\usepackage{xcolor}
\usepackage{float}
\usepackage{graphicx}
\usepackage{csquotes}
\usepackage{amsmath}
\usepackage{subcaption}
\usepackage{comment}
\usepackage[breaklinks,hidelinks]{hyperref}
\usepackage[capitalise]{cleveref}

\def\BibTeX{{\rm B\kern-.05em{\sc i\kern-.025em b}\kern-.08em
    T\kern-.1667em\lower.7ex\hbox{E}\kern-.125emX}}

\AtBeginDocument{%
  \providecommand\BibTeX{{%
    \normalfont B\kern-0.5em{\scshape i\kern-0.25em b}\kern-0.8em\TeX}}}
\usepackage{amsmath}
\usepackage{authblk}
\providecommand\iff{\DOTSB\;\Longleftrightarrow\;}

\newcommand{\loadshedder}{Load Shedder}
\newcommand{\todo}[1]{\noindent{\color{red}{\bf \fbox{TODO}}{\it#1}}}

\begin{document}

\title{Utility-Aware Load Shedding for Real-time Video Analytics at the Edge}

\author[1]{Enrique Saurez$^*$\thanks{esaurez@gatech.edu}}
\author[1]{Harshit Gupta$^*$\thanks{harshitg@gatech.edu}}
\author[2]{Henriette Röger\thanks{henriette.roeger@ipvs.uni-stuttgart.de}}
\author[2]{Sukanya Bhowmik\thanks{sukanya.bhowmik@ipvs.uni-stuttgart.de}}
\author[1]{\\Umakishore Ramachandran\thanks{rama@gatech.edu}}
\author[2]{Kurt Rothermel\thanks{kurt.rothermel@ipvs.uni-stuttgart.de}}
\affil[1]{Georgia Institute of Technology}
\affil[2]{University of Stuttgart}

\maketitle
\def\thefootnote{*}\footnotetext{These authors contributed equally to this work.}
\def\thefootnote{}\footnotetext{This work was supported by the German Research Foundation (DFG) under the research grant "PRECEPT II" (RO 1086/19-2 and BH 154/1-2).}
\thispagestyle{plain}
\pagestyle{plain}
\begin{abstract}

Real-time video analytics typically require video frames to be processed by a query to identify objects or activities of interest while adhering to an end-to-end frame processing latency constraint. Such applications impose a continuous and heavy load on backend compute and network infrastructure because of the need to stream and process all video frames. Video data has inherent redundancy and does not always contain an object of interest for a given query. We leverage this property of video streams to propose a lightweight \loadshedder{} that can be deployed on edge servers or on inexpensive edge devices co-located with cameras and drop uninteresting video frames. The proposed \loadshedder{} uses pixel-level color-based features to calculate a utility score for each ingress video frame, which represents the frame's utility toward the query at hand. The \loadshedder{} uses a minimum utility threshold to select interesting frames to send for query processing. Dropping unnecessary frames enables the video analytics query in the backend to meet the end-to-end latency constraint with fewer compute and network resources. To guarantee a bounded end-to-end latency at runtime, we introduce a control loop that monitors the backend load for the given query and dynamically adjusts the utility threshold. Performance evaluations show that the proposed \loadshedder{} selects a large portion of frames containing each object of interest while meeting the end-to-end frame processing latency constraint. Furthermore, the \loadshedder{} does not impose a significant latency overhead when running on edge devices with modest compute resources.

\end{abstract}


\section{Introduction}
Real-time video analytics has been gaining rapid popularity due to its utility in applications such as surveillance\cite{xu2019space}, driving assistance and safety\cite{le2010real}, and factory automation\cite{factory}. Such applications are typically structured as a pipeline of operators with the first operator consuming a stream of video frames from cameras and feeding its output to the next operator, and so on. Each operator executes a piece of the overall application logic, for instance, object detection, classification, activity recognition, etc., and extracts relevant insights from camera frames. We specifically focus on video analytics pipelines with stringent end-to-end latency constraints, such that the extracted insight from video processing could be used to trigger a real-time response, e.g., alert to car driver. The increasing availability of high quality and high frame rate cameras put significant pressure on the backend compute and networking resources. Although the use of edge resources for running geo-distributed video analytics has been proposed to minimize backhaul bandwidth usage \cite{ananthanarayanan2017real}, the resource capacity of edge sites is typically limited due to space and power constraints \cite{ramachandran2021case}. Oftentimes complex operators like object detection pose heavy compute requirements, such as access to a GPU, which imposes limitations on the number of cameras that can be served at a given edge site.
\par Video streams possess two key characteristics that enable serving more number of cameras with limited resources. Firstly, the appearance of the object-of-interest for a given analytics query is not frequent\cite{nossdav}, implying that a large fraction of camera frames do not contain useful information. Secondly, when an object-of-interest exists in a video stream, due to the high frame rate of cameras it usually is present in multiple frames. Dropping a small portion of the frames that an object-of-interest appear in does not affect the overall fidelity of the results. These characteristics motivate the use of load shedding techniques to shed irrelevant frames, to reduce the workload on the application pipeline. Previous work in load shedding has focused on using linear selectivities \cite{Tatbul1,Tatbul2} and work with structured queries and data \cite{espice, pspice, bo:2020}. Such techniques haven't been explored for content-based shedding of unstructured data such as video. Previous work in early-discard of video frames either require expensive hardware for feature extraction~\cite{filterforward, zhang2020fast} or do not tune the filtering parameters according to the processing load on the backend video analytics pipeline~\cite{reducto}.
\par In this work, we present a lightweight load shedding technique that uses a per-query content-based utility function to determine if a frame should be shed. The utility of a frame is calculated as a function of its color distribution. Each query undergoes a learning phase during which the utility function is built. The \loadshedder{} receives all ingress frames and it drops those whose utility is below the baseline utility threshold. Due to inherent variations in video streams' contents, the utility threshold needs to be dynamically tuned so that the load on backend analytics pipeline is within manageable levels, and the end-to-end processing latency constraint of the query is continuously met. 
%
The \loadshedder{} includes a feedback control-loop that dynamically updates the utility threshold based on the current load on the later stages of the video processing pipeline.
%
This feedback from the later stages ensures that the overall query processing pipeline functions correctly despite differences in the content of the video stream compared to the training set.
We incorporate the proposed load shedding technique on a video analytics platform and perform extensive evaluations with real-world analytics queries and video datasets. Our contributions can be summarized as follows:
\begin{enumerate}
\item A workflow for building the per-frame utility function given a query and a labeled training data set. The utility function should be lightweight, i.e., it should be able to process a high rate of ingress frames without imposing significant latency overhead.
\item  A control loop that dynamically tunes the utility threshold based on the current load on the query operators. The objective of the control loop is to keep the end-to-end latency under a query-specific bound. 
\item Performance evaluation of the proposed load shedding approach to demonstrate its efficacy.
\end{enumerate}

\par The rest of the paper is structured as follows. \cref{sec:background} discusses the context of this work and \cref{sec:relwork} presents the related work
in the space of load shedding for video analytics queries. \cref{sec:load_shedding_algo} presents the system architecture and the load shedding approach in detail, including the training phase and the dynamic tuning of utility thresholds. \cref{sec:evals} presents the design and results of experimental evaluation of the proposed approach. \cref{sec:discussion} presents a discussion of relevant questions stemming from this research. \cref{sec:conclusion} ends the paper with conclusion and a discussion of future work.
\section{Background and Problem Statement}
\label{sec:background}
We describe the context in which our proposed approach is to be employed, followed by a formal definition of the problem statement.

\subsection{Context}
\label{sec:context}
\begin{figure}[ht]
\centering
\includegraphics[width=0.9\columnwidth]{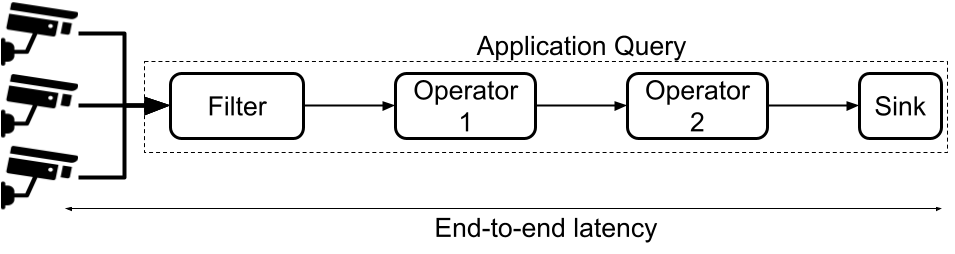}
\caption{Query model.}
\label{fig:query_model}
\end{figure}

\par We target real-time video analytics queries that perform detection of target objects of a specific color. Such queries are common in the domains of surveillance (e.g, tracking red cars in response to an AMBER alert \cite{pillai2020vehicle}), traffic control  (e.g., detecting if an emergency vehicle is stuck in traffic \cite{roy2019emergency}), search and rescue (e.g., locating humans in open water using drones \cite{varga2022seadronessee}), etc. Such queries typically process multiple frames containing a given target object (e.g., a suspicious red car in first example) to extract insights about the object (such as its direction of motion, or which street it went to from an intersection). The query could be designed to process frames containing target objects of a single color (e.g., red suspicious vehicle), frames containing at least one object from the target colors (e.g., white ambulance or red fire truck) or frames containing objects of all target colors.
\par \cref{fig:query_model} shows the query model at a high level, which is composed of three components: a filter, a sequence of one or more video processing operators (e.g., Deep Neural Network based object detection), and a sink. The filter discards frames with no useful information - for instance, filtering out frames with contiguous groups of pixels (blobs) of a certain color larger than a certain size. The video processing operators form the core of the query logic. Finally, the sink analyzes the output labels of the video processing operators and sends  information to the relevant parties. Depending on the dynamic content of video streams, the fraction of frames processed by the video processing operators (e.g., heavyweight DNNs) and the sink varies with time. This variation in processing load causes significant variation in end-to-end latency of the system as well. To ensure real-time response, the queries have a constraint on end-to-end latency of processing a frame containing a target object. The end-to-end processing latency of a frame is the total time taken between the generation of the frame by the camera to the time when it is executed completely by the application query (including communication delay).
\begin{figure}[htp]
\subfloat[Scenario with edge server hosting the Load Shedder along with application query. Compute on the edge is the bottleneck resource.]{%
  \includegraphics[clip,width=\columnwidth]{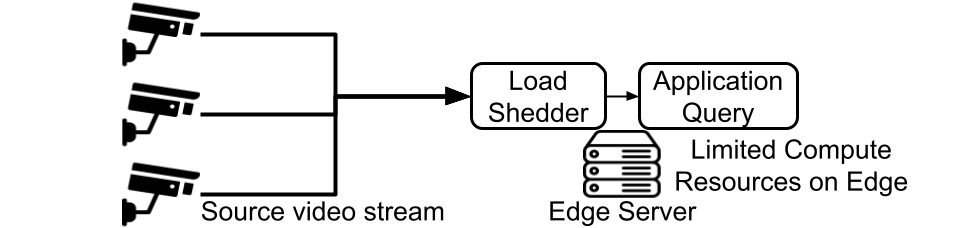}%
}

\subfloat[Scenario with edge server hosting the Load Shedder and cloud hosting the application query. Network bandwidth between edge and cloud is the bottleneck resource.]{%
  \includegraphics[clip,width=\columnwidth]{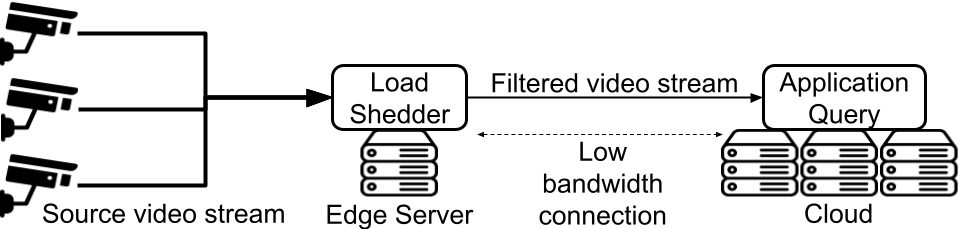}%
}

\subfloat[Scenario with the camera hosting the Load Shedder and cloud hosting the application query. Network bandwidth between cameras and cloud is the bottleneck resource.]{%
  \includegraphics[clip,width=\columnwidth]{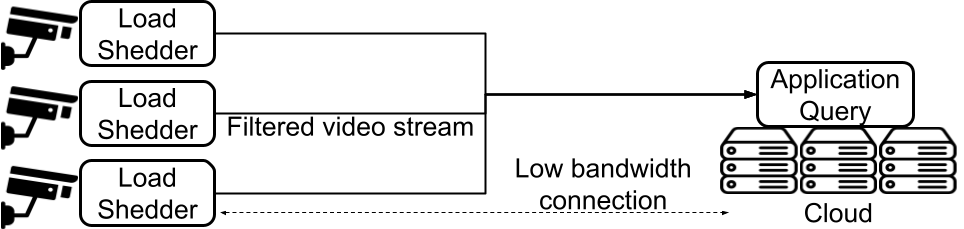}%
}
\caption{Three deployment scenarios for a connected camera system in which the proposed load shedding approach can be employed.}
\label{fig:deployment_scenarios}
\end{figure}
\par We consider a connected camera deployment which could (but not necessarily) be assisted by an edge compute node available in proximity. We assume cameras to contain limited compute capability for running background subtraction and feature extraction (to be discussed in \cref{sec:building_model}). Cameras send the foreground of frames along with the associated features downstream. The possible scenarios of the deployment of downstream components, i.e., the \loadshedder{} and Application Query, are shown in \cref{fig:deployment_scenarios}. In each such scenario, either the compute resource on the edge or the network bandwidth between edge and cloud or between camera and cloud is the bottleneck resource, whose over-utilization results in excessive queuing of video frames, eventually resulting in violation of the end-to-end processing latency constraint. Hence, the \loadshedder{}'s effective operation is crucial to make sure that the bottleneck resource is not overloaded. The objective of the \loadshedder{} is to maximize the fraction of frames containing each target object that are sent downstream to the application query after shedding. The \loadshedder{} optimizes this objective with the constraint of end-to-end processing latency being below the query-specific bound.
 
 \subsection{Problem Statement} 
 We now formally define the problem statement to be solved by the \loadshedder{} by mathematically expressing the objective function and latency constraint. 
 We define video streams as a continuous sequence of frames. Therefore, the source video stream generated from a camera is denoted by  $V=[f_1, f_2, \cdots, f_m]$.  We define an application query as $Q = [ q_1, \cdots, q_n ]$, where $q_i$ denotes a video processing operator (including filtering), which reads input from the output of $q_{i-1}$ and feeds into the input of $q_{i+1}$. The first operator in a query always reads the combined video stream coming from the multiple cameras it is serving, as shown in \cref{fig:query_model}.  We denote the output stream of operator $q_i$ with input stream $v$ as $q_i \left( v \right)$. We define the set of target objects detected by a query $Q$ in video $V$ as $T_Q \left( V \right)$ as in \cref{eq_target}. We represent frame $f$ containing target object $o$ by the relation $o \in f$.
 \begin{equation}
 T_Q \left( V \right) = \{ \text{target objects in } V \text{ detected by } Q \}
 \label{eq_target}
 \end{equation}
 
Now, introducing the \loadshedder{} component into the video query $Q$, we construct a query $Q'$ of the form $[q_0, q_1, \cdots, q_n]$, where $q_0$ is the \loadshedder{} component, also denoted by $LS$. The \textit{Quality of Result (QoR)} metric we use in this work is designed to measure the number of frames for each target object that are sent downstream by the \loadshedder{} $LS$ in query $Q'$, i.e., belonging to the output stream $LS \left( V \right)$. We define the per-target-object QoR for target object $o$ in \cref{per_obj_qor}.  QoR metric has a value between 0 and 1. 
 \begin{equation}
QoR_Q \left( o, LS, V \right) = \dfrac{|f \in LS\left( V \right) : o \in f|}{|f \in V : o \in f|}
\label{per_obj_qor} 
 \end{equation}
 The overall QoR metric for query $Q$ with the \loadshedder{} $LS$ against source video $V$ encompassing all target objects of interest is defined in \cref{qor}. This metric calculates the average per-object QoR metric over all target objects. Thus, this metric quantitatively measures the aggregate performance of the \loadshedder{} for a given source video.
 \begin{equation}
 QoR_Q \left( LS, V \right) = \dfrac{\sum_{o \in T_Q \left( V \right)} QoR_Q \left( o, LS, V \right)}{|T_Q \left( V \right)|}
\label{qor}
\end{equation}

\par For a given video stream $V = [  f_1, f_2, \cdots , f_m ]$, we define the end-to-end delay experienced by frame $f$ of video $V$ when processed by query $Q'$ as $E2E_{V, Q'} \left( f \right)$ which is expressed as shown in \cref{e2edelay}, where $q_k$ is the last operator in $Q'$ that processes frame $f$.
 
\begin{equation}
E2E_{V, Q'} \left( f \right) = \sum_{i=0}^{k} queue \left( q_i, f \right) + exec \left( q_i, f \right) 
\label{e2edelay}
\end{equation}
 
The  objective is to maximize the Quality of Result (QoR), 
while dropping frames such that the given end-to-end latency bound $LB$ is met.  More formally, the objective is defined as follows.
\begin{equation}
\begin{aligned}
Maximize \quad  & \ QoR_Q \left(LS, V \right)	\\
\textrm{s.t.} \quad & \ E2E_{V, Q'} \left( f \right) \le LB \quad \forall~ f \in V
\end{aligned}
\end{equation}

\section{Related Work}
\label{sec:relwork}
\par \noindent \textbf{Resource Management in Online Video Analytics}: Live video analytics is an extremely compute and network resource intensive application. 
Prior art has adopted two main approaches for managing the high resource requirements of this application - tuning the configuration of the input stream and video processing operators (e.g., frame rate, resolution, etc.) and early discard of uninteresting frames. Online adaptation of camera streams and video processing operators has been explored in several prior works \cite{videostorm,chameleon,soudain}. These works target scenarios where end-to-end latency of the order of sew seconds can be tolerated. Their primary objective is to optimize cost of resources needed for supporting all video streams while also maintaining high accuracy. Although these works differ from our proposed approach in terms of objectives, their technique can be adopted to work complementary to our proposed approach.
\par \noindent \textbf{Early-Discard Filtering of Video}: Multiple approaches have been taken for performing early discard of unnecessary video frames so that compute and network costs of streaming and processing them can be avoided. Glimpse \cite{glimpse} proposed a hardware-software add-on that could use low-powered coarse-grained vision modalities to filter out frames irrelevant to the query at hand. The system uses specialized hardware for motion detection, temperature measurement, etc., which are not available on the cameras in our scenarios. Zhang, et al., \cite{fastfiltering} use a multi-stage pipeline of operators to determine whether a frame is relevant to a query, wherein the operators are implemented using GPUs and impose significant latency overhead to the video processing pipeline's performance. FilterForward \cite{filterforward} consists of a base DNN whose intermediate layers' outputs are used by a per-application binary classifier that decides whether to filter out a frame or not. However, in that system, the execution latency of the base model itself is very high (>200ms) and the cost of performing early discard amortizes only when running at least 4 concurrent application queries per base model. The EarlyDiscard strategy proposed in Wang, et al., \cite{drones} uses an inexpensive DNN to perform filtering of frames. However, there is no way to tune the filter so as to ensure end-to-end latency guarantee of video processing. Reducto \cite{reducto} makes use of difference in low-level visual features (such as pixels, edges, etc.) across consecutive frames to determine if processing the new frame would result in a difference in the query's result. However, it operates at 1-second granularity, meaning that it cannot guarantee end-to-end latency less than that. Additionally, it does not support tuning of the filtering based on load experienced by backend application query.
\par In summary, previous work on early discard of frames either use expensive filters that add significant latency overhead to the video processing pipeline, or do not support the fine-tuning of the filter to ensure end-to-end latency.
\section{Utility-Aware Load-shedding}
\label{sec:load_shedding_algo}
In this section, we present a high-level architecture of the proposed load shedding system, highlighting the interactions between different components. Then we describe the design and functionality of each individual component in isolation.
\newcommand{\trainingset}{$\mathcal{D}$ }

\subsection{System Architecture}\label{sec:design}
\begin{figure}[ht]
\centering
\includegraphics[width=0.9\columnwidth]{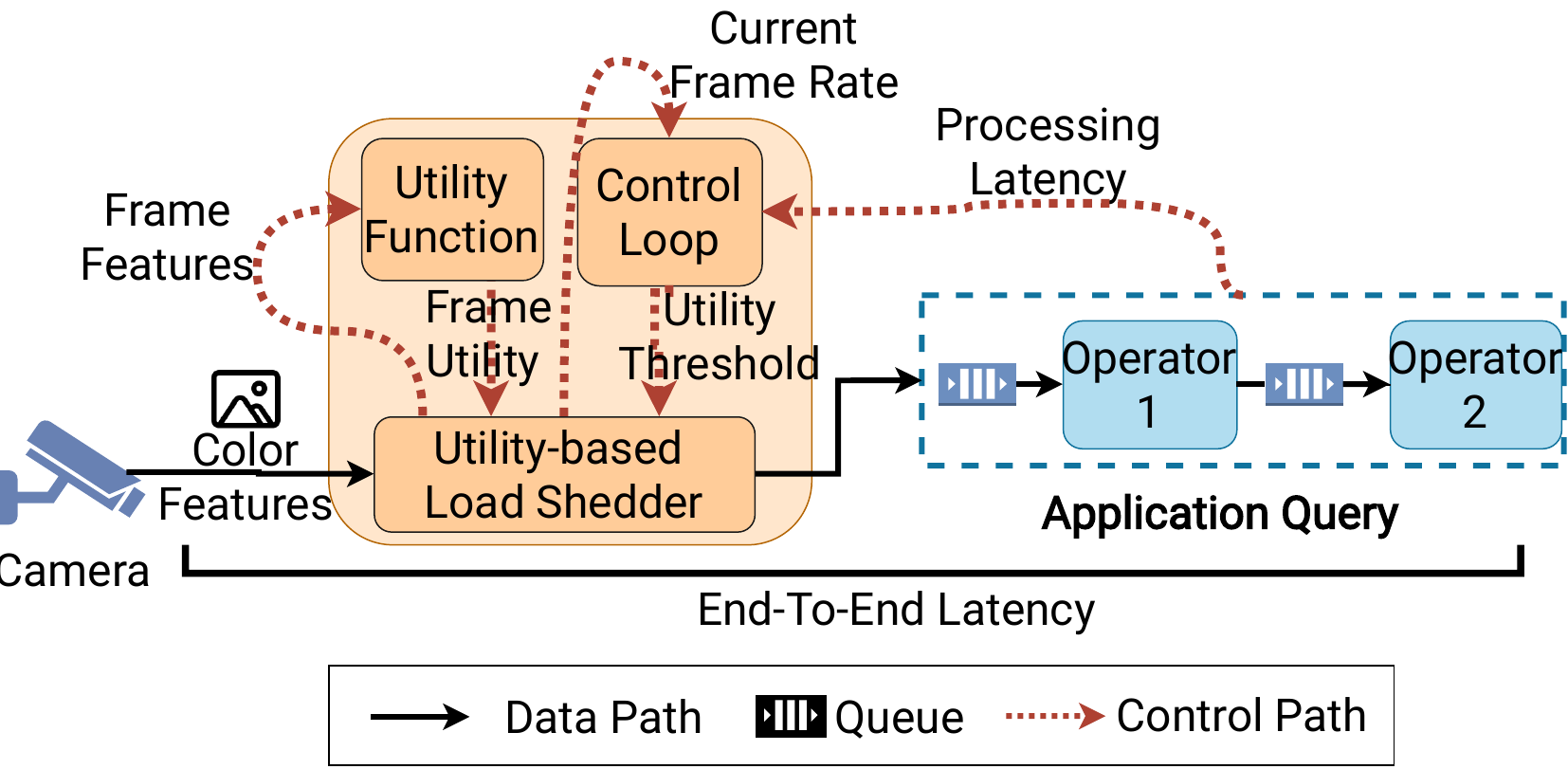}
\caption{Architecture of the proposed load-shedding approach.}
\label{fig:arch}
\end{figure}

We extend  real-time video analytics systems which are equipped with edge-computing capability (referring to both edge sites and compute capability co-located with cameras) with our proposed load shedding system that sheds a portion of the input frames to maintain the given latency bound ($LB$) during overload. Figure \ref{fig:arch} shows the main components involved in performing the aforementioned tasks and their interactions---the \loadshedder{}, the application query, and the Control Loop. Each operator (including \loadshedder{}) reads from an ingress queue which is (populated by the upstream operator or source cameras) and pushes the output data to the egress queue. 
\par The load shedding system must perform two primary tasks---(1) decide when to shed and how much to shed, and (2) decide which frames to shed. We describe the latter task first, for which the \loadshedder{} computes a utility/ importance value for each video frame that 
denotes the probability of the frame being \textit{useful} for the application query. The utility is calculated using the color content of the frame and the application query at hand. The intuition behind the design of the \loadshedder{} is to discard video frames that have a low probability of being useful for the given application query (see \cref{sec:building_model}). The \loadshedder{} can be dynamically configured with a utility threshold, such that it drops frames with utility less than the threshold.
\par The Control Loop component  dynamically computes the utility threshold and updates the \loadshedder{}. The utility threshold calculation is done so as to sustain the current load on the application query, while meeting the end-to-end latency bound ($LB$) (see \cref{sec:control-loop}). More specifically, the Control Loop component monitors the queue lengths of each 
operator in the 
query, and estimates the current observed end-to-end latency. Based on the observed end-to-end latency and the 
query's latency requirement, the Control Loop computes the fraction of frames that should be dropped by the \loadshedder{}, which is then transformed into a utility threshold (see \cref{sec:util_threshold}). Transforming the desired frame drop rate to a utility threshold is done on the basis of the utility distribution of frames observed in the past.

\par The subsequent sections describe the per-frame utility calculation function (exploration of color features for its design), the calculation of utility threshold for a given target drop rate, and the control loop in more detail.



\begin{table}
\begin{center}
\begin{tabular}{|| c | c ||} 
 \hline
 Notation & Description \\ [0.5ex] 
 \hline\hline
 \trainingset & Training data set \\ 
 \hline
 f & Frame \\ 
 \hline
 l & Label $\in \{ 0, 1\}$\\ 
 \hline
 C & Color as a range of Hue\\
 \hline
 $HF_C \left( f \right)$ & Hue Fraction of frame $f$ for color $C$\\ 
 \hline
 $PF_C^{\left( i,j \right)} \left( f \right)$ & Pixel Fraction of frame $f$ for color $C$ \\
 & and saturation-value bin $\left( i,j\right)$\\
 \hline
 $M_{C, +ve}^{\left( i,j \right)}$ & Correlation metric of saturation-value  \\
 & bin $\left( i,j\right)$ of color $C$ with +ve frames \\
 \hline
 $U_C \left( f \right)$ & Utility value of frame $f$ for color $C$\\ 
 \hline
\end{tabular}
\caption{Notations used in the formulation.}
\end{center}
\end{table}

\subsection{Building the utility function}
\label{sec:building_model}
The \loadshedder{} makes the decision of whether to discard a frame or not based on each frame's utility value - which is calculated by the utility function. It takes as input the frame's color-based features and computes a scalar utility value. 

\subsubsection{Hue-Saturation-Value Color Model}
\begin{figure}[ht]
\centering
\includegraphics[width=0.7\columnwidth]{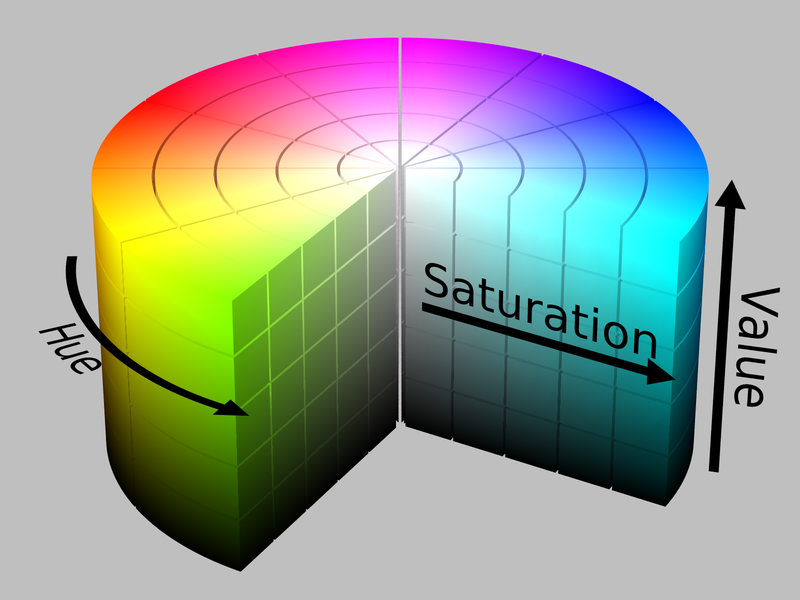}
\caption{
HSV Color Model \cite{hsv_wiki}. Hue represents the color itself (or the dominant wavelength), Saturation represents the brilliance and intensity of the color, while Value defines the lightness or darkness of the color. The ranges of Hue, Saturation and Value we use are $[0, 180 )$, $[0, 256)$, and $[0,256)$ respectively.}
\label{fig:hsv}
\end{figure}

We use the Hue-Saturation-Value (HSV) model for representing the color of pixels, which represents how colors appear under light (\cref{fig:hsv}). The HSV model is widely used in computer vision applications over the Red-Green-Blue (RGB) model because HSV separates the color information from intensity information (which helps deal with situations like lighting changes or shadows). We use the HSV triplet of each pixel to calculate the distribution of colors in each frame, which forms the input feature set for the utility function (described in detail in subsequent sections). The developer of the Application Query specifies the color of the target objects in terms of a hue range $C$ as input to the \loadshedder{}. For instance, the color red is represented using the hue ranges $[0,10) \cup [170,180)$. 

\subsubsection{Training data}
To train the utility function, we use a labeled stream of videos from multiple cameras as training data set. Each element of the training data set \trainingset is of the form $\left(f, l\right)$, where $f$ represents the frame and $l$ represents the label. Each frame $f$ is described by a list of pixels, with each pixel being represented by a triplet $\left( h, s, v \right)$. The $h, s, v$ elements of a pixel's notation denote the pixel's Hue, Saturation and Value fields respectively. The label $l$ tells whether the frame contains an object of interest, i.e., whether the frame was a match for the given query. Henceforth, we use the term \textit{positive} to denote frames which contain one or more target objects, and \textit{negative} to denote frames which do not contain them. 

\par The goal of the utility function is to be able to separate the computed utilities of positive and negative frames, such that using a utility threshold would result in effective load shedding. In the remaining part of this section, we first outline our observations from the training data on how to separate positive and negative frames using the HSV model. Based on these observations, we build our utility function.

\subsubsection{Hue as feature for separating positive and negative frames}
\begin{figure*}[t!]
    \centering
    \begin{subfigure}[t]{0.45\textwidth}
        \includegraphics[width=\textwidth]{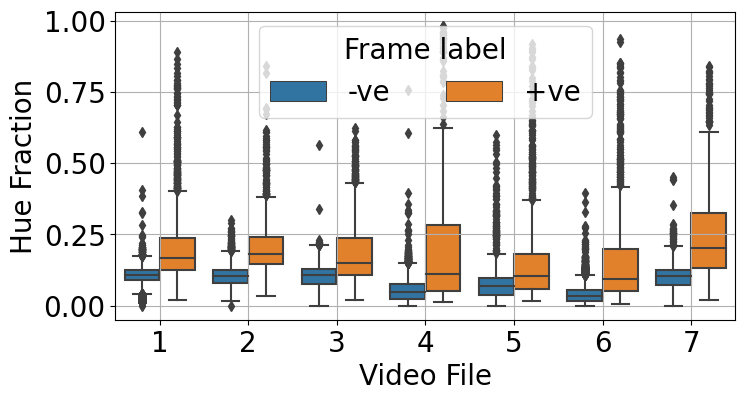}
        \caption{Distribution of Hue Fraction of color RED over videos in the training dataset.}
        \label{fig:hf_all_vids}
    \end{subfigure}%
    \hfill 
    \begin{subfigure}[t]{0.45\textwidth}
        \includegraphics[width=\textwidth]{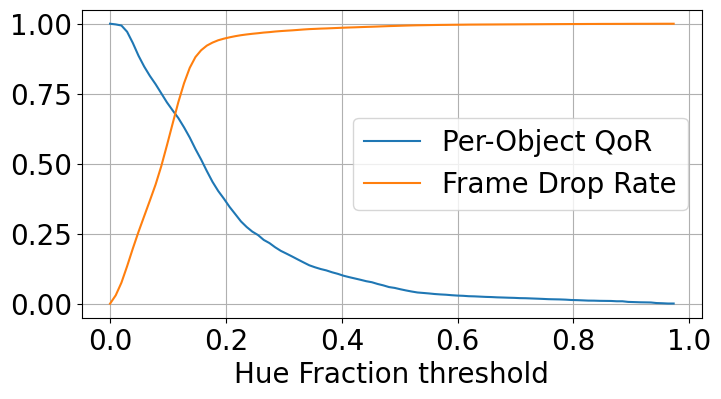}
        \caption{QoR and Drop Rate (both ranging from 0 to 1)  vs. HF threshold for RED color.}
        \label{fig:hf_qor}
    \end{subfigure}
    \caption{Results of Hue Fraction as a feature for discriminating between positive and negative frames. \cref{fig:hf_all_vids} shows that Hue Fraction of negative frames has significant overlap with positive frames. \cref{fig:hf_qor} shows that high frame drop rate cannot be achieved without a significant drop in QoR metric.}
    \label{fig:hf_eval}
\end{figure*}

We first explore the use of the Hue field of pixels to calculate whether the given frame contains a target object of the given color. We do so by computing a metric \textit{Hue Fraction} for the color $C$, denoted by $HF_C$. 
\begin{equation}
HF_C \left( f \right) = \dfrac{| \text{pixel }p \in f: hue \left( p \right) \in C |}{|\text{pixel }p \in f|}
\end{equation}
A higher Hue Fraction would imply higher likelihood for the frame to contain a target object of a given color. Hence, a threshold-based approach on $HF_C$ is a candidate for the \loadshedder{}. However, our analysis of hue fractions of frames in our dataset showed that the distribution of $HF_C$ for negative frames overlaps significantly with positive frames across all videos. For instance, \cref{fig:hf_all_vids} shows the hue fractions of the color Red of all frames across all videos in the dataset; it can be seen that the hue fraction distribution of positive frames overlaps with negative frames. This overlap would prevent a threshold-based approach on the hue fraction from effectively differentiating between positive and negative frames. This effect is captured in \cref{fig:hf_qor}, where the per-object QoR metric drops steeply with hue fraction thresholds without achieving a significant frame drop rate. We posit that this overlap in distribution is because both positive and negative frames contain red-colored pixels; however, the saturation and value fields of those pixels in the positive and negative frames would have distinctly discernible distributions for the two. 

\subsubsection{Using Saturation and Value fields for differentiating frames}
In order to separate positive frames from negative ones, we analyze the distribution of Saturation and Value fields for Red pixels across videos. We discretize the range of pixel saturation and value into bins of size $s$ and $v$ respectively. We define two functions $sat\_bin$ and $val\_bin$ that map a pixel's saturation $sat\left( p \right)$ and value $val\left( p \right)$ respectively to their corresponding bins.
\begin{equation}
sat\_bin \left( p \right) = i \iff i \cdot s \leq sat \left( p \right) < \left( i+1 \right) \cdot s
\end{equation}
\begin{equation}
val\_bin \left( p \right) = j \iff j \cdot v \leq val \left( p \right) < \left( j+1 \right) \cdot v
\end{equation}

We leverage the division of the saturation and value range into bins to transform each frame $f$ into  a two-dimensional matrix for a given color $C$, such that each element of the 2D matrix denotes the fraction of pixels belonging to a particular saturation and value bin.
\begin{equation}
\resizebox{.9\hsize}{!}{$PF_C \left( f \right) = 
\begin{pmatrix}
PF_C^{\left( 0,0 \right)}\left( f \right) & PF_C^{\left( 0,1 \right)}\left( f \right) & \cdots & PF_C^{\left( 0,n \right)}\left( f \right) \\
PF_C^{\left( 1,0 \right)}\left( f \right) & PF_C^{\left( 1,1 \right)}\left( f \right) & \cdots & PF_C^{\left( 1,n \right)}\left( f \right) \\
\vdots  & \vdots  & \ddots & \vdots  \\
PF_C^{\left( m,0 \right)}\left( f \right) & PF_C^{\left( m,1 \right)}\left( f \right) & \cdots & PF_C^{\left( m,n \right)}\left( f \right) 
\end{pmatrix}$}
\end{equation}
Where $PF_C^{\left( i,j \right)}$ is the fraction of pixels whose hue falls in the range $C$ and their saturation and value fall into the $i^{th}$ and $j^{th}$ bin respectively. The number of bins for saturation and value are $B_S$ and $B_V$ respectively.

\begin{equation}
\resizebox{.9\hsize}{!}{$PF_C^{\left( i,j \right)} \left( f \right) = \dfrac{| \text{pixel } p \in f : hue \left( p \right) \in C \land in\_bin(p,i,j)  |}{| \text{pixel } p \in f : hue \left( p \right) \in C|}$}
\end{equation}

\begin{equation}
in\_bin(p,i,j) = \left( bin_{sat} \left( p \right) = i \land bin_{val} \left( p \right) = j \right)
\end{equation}

\begin{figure}[ht]
\centering
\includegraphics[width=\columnwidth]{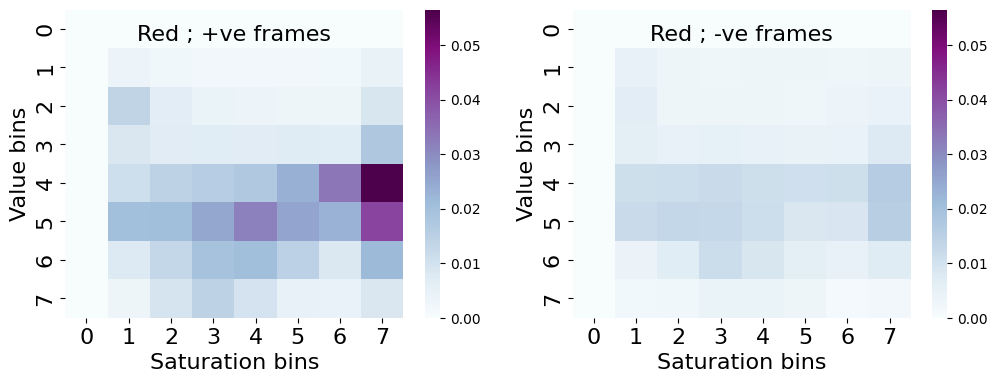}
\caption{Distribution of Saturation and Value fields for positive ($M_{C, true}$) and negative frames ($M_{C, false}$) across all videos in the dataset. Bins with high saturation are better differentiators of positive frames.}
\label{fig:sat_val_bins_red}
\end{figure}

Using the per-frame distribution of pixels in various sat\-u\-ration-value bins, we quantify how useful each such bin is in classifying a frame as positive or negative. We compute a metric for each saturation-value bin using the pixel distribution of positive and negative frames that denotes the correlation of the bin with the particular label (positive or negative) for a given frame. 

\begin{equation}
M_{C, +ve}^{\left( i,j \right)} = \text{AVG } PF_C^{\left( i,j \right)} \left( f \right) \forall \left( f, 1 \right) \in \mathcal{D}
\label{eq:sat_val_bin_util_positive}
\end{equation}

\begin{equation}
M_{C, -ve}^{\left( i,j \right)} = \text{AVG } PF_C^{\left( i,j \right)} \left( f \right) \forall \left( f, 0 \right) \in \mathcal{D}
\label{eq:sat_val_bin_util_negative}
\end{equation}

The distribution of above utility for the color Red, i.e., $M_{C, true}$ and $M_{C, false}$ is shown in \cref{fig:sat_val_bins_red}. The bins with higher saturation value are much stronger indicators of whether a frame is positive.

\subsubsection{Computing per-frame utility}
We use the utility of each saturation-value bin (from \cref{eq:sat_val_bin_util_positive} and \cref{eq:sat_val_bin_util_negative}) to compute the utility value for a frame. The utility is a weighted sum of the $M_{C,+ve}^{\left( i,j \right)}$ value for each saturation-value bin, weighted by the pixel fraction of the frame in that bin.
\begin{equation}
U_C \left( f \right) = M_{C,+ve}^{\left( i,j \right)} \cdot PF_C^{\left( i,j \right)} \left( f \right) 
\end{equation}

\begin{figure}[ht]
\centering
\includegraphics[width=0.95\columnwidth]{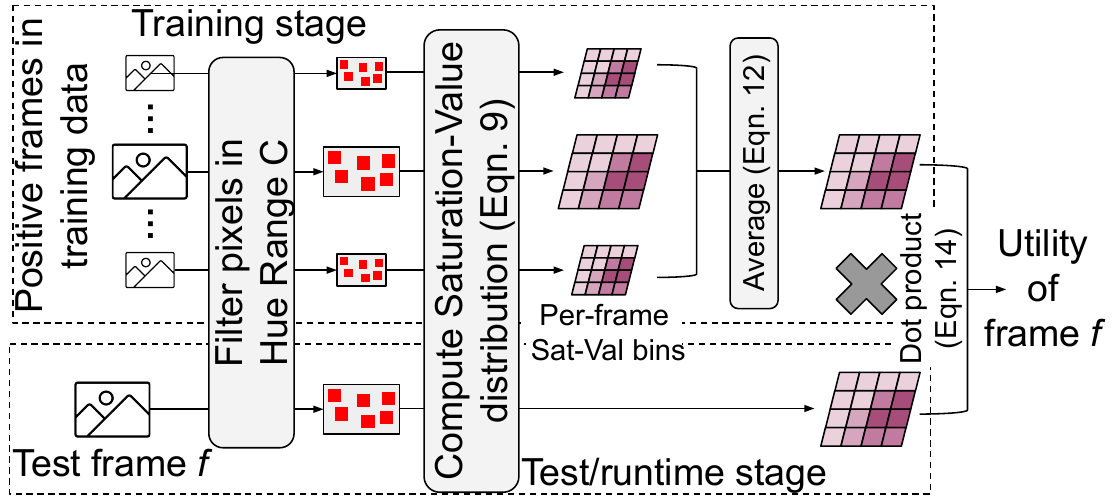}
\caption{A schematic description of the proposed utility-based load shedding approach. The figure shows both training and testing/runtime stages, and the steps involved for both. Equations referring to specific computations in the figure have been referenced.}
\label{fig:util_model_walkthrough}
\end{figure}
\cref{fig:util_model_walkthrough} schematically describes the approach of building the utility function using training data and using the function to calculate utility value for frames at runtime.

\subsubsection{Computing per-frame utility for composite color queries}
Computing the utility of a frame for composite queries requires using the utility function computed for the component colors. 
\begin{equation}
U_{C_1 \lor C_2} \left( f \right) = max \left( \overline{U_{C_1}} \left( f \right), \overline{U_{C_2}} \left( f \right)  \right)
\end{equation}
$\overline{U_{C_1}}$ represents the utility function for color $C_1$ normalized over the training data, such that the maximum utility is 1.0. Normalization of per-color utility functions allows their effective composition. Similarly for computing the utility of a query looking for both colors $C_1$ \textbf{AND} $C_2$ in a frame, we use the minimum of $\overline{U_{C_1}}$ and $\overline{U_{C_2}}$ as the composite utility.

\subsection{Computing Utility Threshold}
\label{sec:util_threshold}
The utility threshold to be used at any given point of time by the \loadshedder{} is computed using the target frame drop rate that is currently desired. This mapping is learned using the distribution of utility values for a set of frames $\mathcal{H}$ in recent history. Using the utility function $U_C$, we build a cumulative distribution function (CDF) of utility values over frames in the history $\mathcal{H}$, as shown in equation \cref{eq:cdf}.
\begin{equation}\label{eq:cdf}
CDF \left( u \right) = \dfrac{|\{ f : U_C \left( f \right) \leq u \, \forall \left( f, l \right) \in \mathcal{H} \}|}{|\mathcal{H}|}
\end{equation}
Note that using a CDF-based approach allows incorporating utility values of more recent frames into $\mathcal{H}$ to update the CDF with changing video content. Initially, the training data set $\mathcal{D}$ itself can be used as the set $\mathcal{H}$ of historical frames. To determine the utility threshold for a given target drop rate $r$, we use the inverse of the $CDF$ function iteratively to compute the minimum utility value $u_{th}$ such that 
\begin{equation}
CDF \left( u_{th} \right) \geq r.
\end{equation}
The intuition behind this approach is that for the utility threshold $u_{th}$, the \loadshedder{} will drop a fraction $r$ from $\mathcal{H}$. Since the utility distribution of recent historical frames is expected to be similar to new frames in the near term, the \loadshedder{} is expected to drop $r$ fraction of new incoming frames. However, the observed frame drop rate of new incoming frames might not equal the target drop rate $r$ because it's transformation to utility threshold depends entirely on the distribution of frame utilities in $\mathcal{H}$. Both the target drop rate and observed drop rate have a range from 0 to 1.


\subsection{Design of Control Loop}\label{sec:control-loop}
\par The utility and threshold calculation described in \cref{sec:design} are used by a \loadshedder{} to control the end-to-end (E2E) latency of the execution of a video processing query. The \loadshedder{} requires a control loop (as shown in \cref{fig:arch}) to define the target drop rate to keep the E2E latency within the specified bound.

\subsubsection{Control Loop mechanisms}

\par The developer of the video query defines the required E2E latency, 
which guides the execution of the \loadshedder{}. Besides this requirement, the \loadshedder{} uses as inputs both the frames per second (FPS), the (current) processing latency of backend query execution, and the (current) network latency between camera and \loadshedder{}, and between \loadshedder{} and the backend running the query. The \loadshedder{} uses two main mechanisms to control the end-to-end latency: admission control and dynamic queue sizing. These mechanisms differ primarily in how quickly they adapt the target drop rate based on changes in backend execution load.

\par \noindent \textbf{Admission Control. }The admission control decides which frames are considered for further processing. To do so, it monitors the execution and queuing delays of a frame for all operators of the Application Query and thereby computes the processing latency of each frame by the query. It uses the average perceived query processing latency ($proc_Q$) to calculate the currently supported throughput (ST) by the backend query as:

\begin{equation}
\label{eq:supported_tput}
ST = \dfrac{1}{proc_Q}
\end{equation}

\par The ST is then compared against the FPS (frame per second) coming into the \loadshedder{} to calculate the required target drop rate as follows:

\begin{equation}
\label{eq:target_drop_ratio}
Target~Drop~Rate = max(0, 1-\dfrac{ST}{FPS} ),
\end{equation}

\noindent which prescribes 
the fraction of frames that need to be dropped to match the ingress frame rate going into the backend query with the available processing throughput, such that the system is stable. As described in \cref{sec:util_threshold}, we transform the target drop rate to a utility threshold by using the utility distribution of historical frames, to filter new ingress frames based on their utility value.

\par \noindent \textbf{Dynamic Queue Sizing. } Internally, the \loadshedder{} (as shown in \cref{fig:arch}) also manages a queue that it uses to ensure that the the end-to-end latency of any frame accepted by admission control is met.
The expected E2E latency for the $N^{th}$ frame in the \loadshedder{} queue is shown in \cref{expected_e2e}. $net_{cam,LS}$ and $net_{LS,Q}$ represent the average of continuously monitored network latencies between cameras and \loadshedder{} and between \loadshedder{} and query backend respectively. $proc_{CAM}$ denotes the average latency incurred in processing frames on the camera (including background subtraction, feature extraction, etc.) (analyzed in \cref{sec:breakdown_analysis}). 

\begin{equation}
\begin{split}
Expected~E2E~Latency = (N + 1) \cdot proc_Q +\\ net_{cam,LS} + net_{LS,Q} + proc_{CAM}
\end{split}
\label{expected_e2e}
\end{equation}

\par If one of the component latencies increases, later frames in the \loadshedder{}'s internal queue could violate the E2E latency requirement. Dynamic queue sizing helps reduce the likelihood of a frame with a high utility violating the query's latency requirement.
\par Dynamic queue sizing updates the size of the \loadshedder's queue and drops the frames with the lowest utility when the size is reduced. This design allows the frames with highest utilities to be processed instead of following other policies (e.g., LIFO - that blindly drop older frames). Dynamic queue sizing reacts faster than updates to the utility threshold (that guides admission control) and reduces the likelihood of an E2E latency violation. The queue is always at least of size one to avoid starving the 
downstream operators. 

\par Dynamic queue sizing can also be seen as a second layer of admission control. Even if a new frame has a utility higher than the current threshold, it will be dropped if the queue is full and it has the lowest utility of all frames currently in the queue. 
Similarly, if an incoming new frame has a greater utility than the lowest utility frame that is already in the queue, then the latter will be dropped and the new frame added to the queue.
This queue shedding keeps the latency requirement valid even for new incoming frames.
\par Both mechanisms allow the \loadshedder{} to fine-tune the admission of new frames to extract the most utility out of the video stream while still maintaining the required E2E latency.
\section{Evaluations}
\label{sec:evals}

\par In this section, we present the results of experimental evaluations carried out to test the efficacy of the proposed utility-based load shedding approach. Our experiments are tailored to validate the following hypotheses.
\begin{enumerate}
\item The proposed utility function is able to compute utility value for frames from an unseen video (not in training set) and effectively differentiate between positive and negative frames.
\item The proposed control loop adapts to changing workload pattern and is able to meet application performance requirements without sacrificing QoR metric.
\item The utility value calculation is light-weight and imposes low overhead on edge devices.
\end{enumerate}

\subsection{Data Set}
\par We generated a benchmark of synthetic videos with VisualRoad \cite{visualroad}, which is a benchmark to evaluate video database management systems. VisualRoad uses the autonomous driving CARLA simulator \cite{carla} to generate videos from CCTV cameras located in a realistic city-like environment, including pedestrians, bicycles, different types of vehicles, and all the surroundings (roads and buildings). Additionally, it allows perturbing the locations of cameras (by specifying a seed parameter) and weather conditions, thereby generating a number of different scenarios. 

\par We evaluate our proposed color-based \loadshedder{} and associated control loop with 25 videos from 7 seeds value (3 or 4 videos from each seed value) using sunny weather. Each video represents 15 minutes of a camera video stream facing a road or highway in a city, with a frame rate of 10 fps. Different cameras have different distributions of cars, varying from cars always presents to rarely appearing. In our results, we report metrics for videos that contained a decent number of target objects for the given query.

\subsection{Implementation}

\begin{figure}[ht]
\centering
\includegraphics[width=0.95\columnwidth]{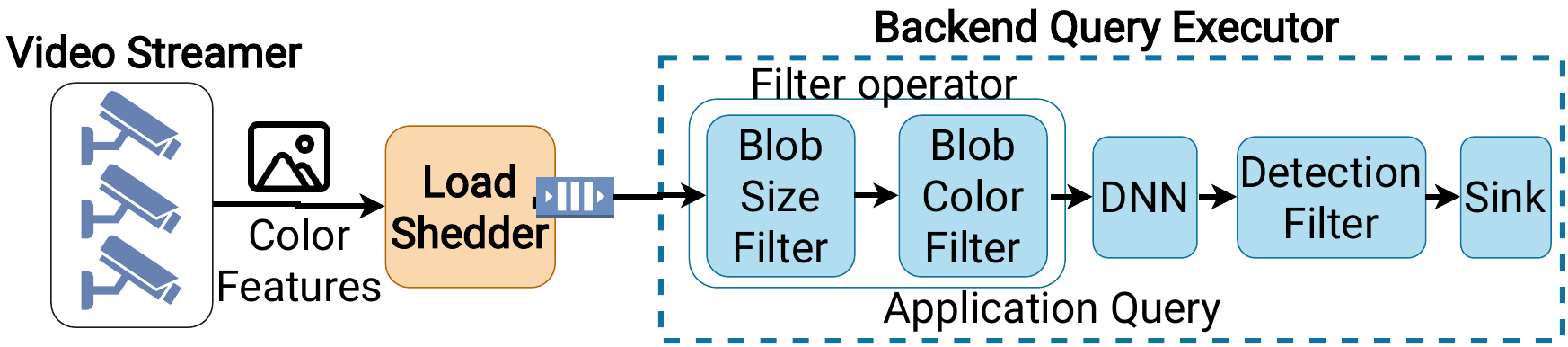}
\caption{The components used in the evaluation setup. The Application Query shows all the components in the considered query.}
\label{fig:evaluation_diagram}
\end{figure}

In order to evaluate the aforementioned hypotheses, we implement the video processing system along with the \loadshedder{}. The system being evaluated has three main components: the Video Streamer, the \loadshedder{}, and the Backend Query executor, as shown in \cref{fig:evaluation_diagram}. All the components exchange messages using the communication library ZeroMQ \cite{zeromq}, with the messages serialized using Cap'n Proto \cite{capnp}. The \loadshedder{} is implemented in Python 3 and all the other components in C++. The Video Streamer reads the video files generated with VisualRoad, performs background subtraction, extracts the color features for each frame (as described in \cref{fig:util_model_walkthrough}) and streams them to the \loadshedder{}. The Video Streamer component is capable of emulating multiple cameras sending their frames' features to the \loadshedder{} by interleaving their frames. Next, the \loadshedder{} implements the utility calculation and load shedding described in \cref{sec:design}. The utility function we use in our evaluations uses 8 bins for both saturation and value, meaning that the bin sizes $s$ and $v$ are equal to 32. Through preliminary experiments (not shown in this paper) we found that these bin sizes offer the best separation of positive from negative frames. Finally, the Backend Query Executor runs the video analytics query to be executed on the video stream. The \loadshedder{} and Backend Query Executor run on a NC6 Virtual Machine on Microsoft Azure with 6 Intel Xeon E5-2690 v3 vCPUs,  56 GiB of RAM and an NVIDIA Tesla K80.


\par There are two additional components to implement the control loop: the Metrics Collector and a Transmission Control Mechanism. The Metrics Collector measures the end-to-end latency in the Backend Query Executor, aggregates it, and forwards it to the \loadshedder{}. The Metrics Collector also monitors the total incoming frames per second. The \loadshedder{} uses this information to define the target drop rate. The Transmission Control Mechanism implements a backpressure algorithm using tokens between the \loadshedder{} and the Backend Query Executor. The latter has a queue that allows the \loadshedder{} to send frames to be processed when a token is freed (i.e., when a frame processing is completed). The tokens allow the \loadshedder{} to control which frames will be processed by the Backend Query Executor when they are running in different machines and balance the trade-off between waiting for a better frame to arrive at the \loadshedder{} and sending the currently best. In addition, the tokens give feedback to the load shedder on when to send the current best next frame to process. If no tokens are remaining, the load shedder should keep analyzing the incoming frames and drop frames with the lower utility if required. On the other hand, if the Backend Query Executor is empty, the load shedder should immediately send something to be processed.  

\subsection{Video queries}
\par Our evaluations consider object detection and tracking queries that need to look at multiple frames of target objects. The model query we use consists of (1) a filter component that groups together spatially adjacent pixels into blobs and drops frames that do not have at least one blob of a certain minimum size, (2) a second filter that ignores frames that do not have a blob(s) of the target object's color, (3) a DNN that performs object detection, (4) a filter that looks for the detected objects' color and label before sending the information to the sink. The object detection DNN we use is \textit{efficientdet-d4} \cite{tan2020efficientdet}, which forms the most computationally intensive components of these components and incurs a bulk of processing latency. As described earlier in \cref{sec:context}, the end-to-end latency of a frame depends on which operators of the query it traverses, which depends on the frame's content. We evaluate simple queries for target objects of a single color (e.g., red) and composite queries for objects of multiple colors (e.g., red or yellow).

\subsection{Performance on Unseen Videos}
We evaluate the performance of the \loadshedder{} on unseen videos using a cross-validation study. We use the video dataset described above and iteratively split it into training and testing set, changing the split in each iteration. In each iteration, we build train the \loadshedder's utility function using the training set and compute the utility and correctness metrics for the test videos.
\subsubsection{Single-color query: Red}
\begin{figure*}[ht]
    \centering
    \begin{subfigure}{0.6\textwidth}
        \includegraphics[width=\textwidth]{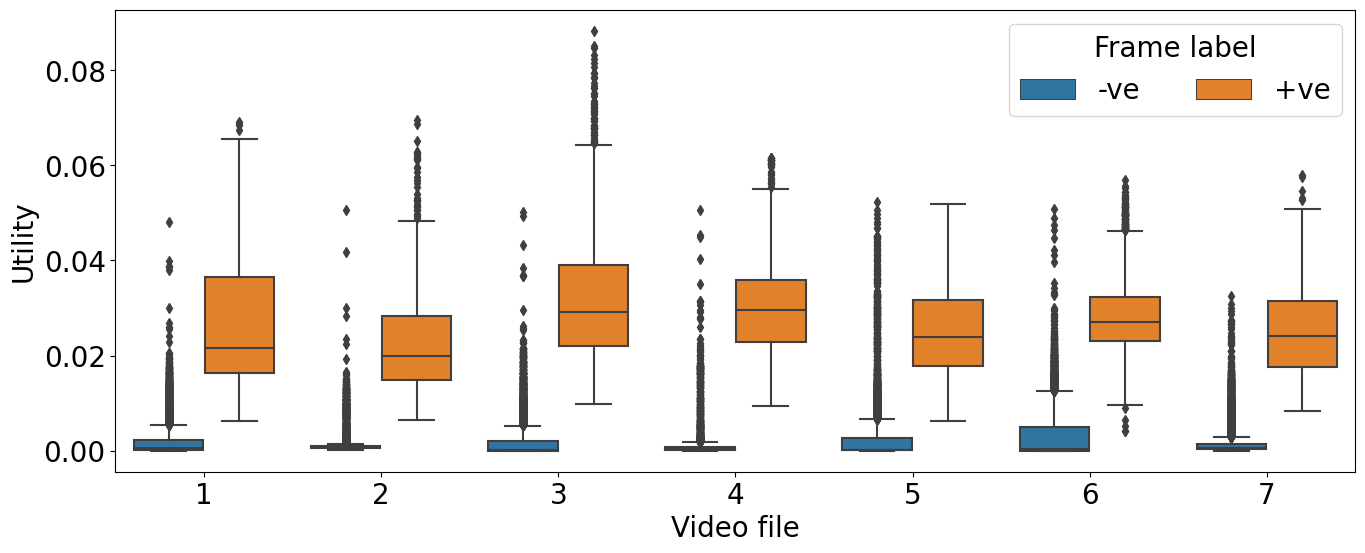}
        \caption{Utility values for positive and negative frames of unseen video frames. For a given video in the figure, the utility function used to compute its frames' utility values is trained using data that does not contain the given video.}
        \label{fig:util_cross_val_red}
    \end{subfigure}
    \hfill
    \begin{subfigure}{0.35\textwidth}
        \includegraphics[width=\textwidth]{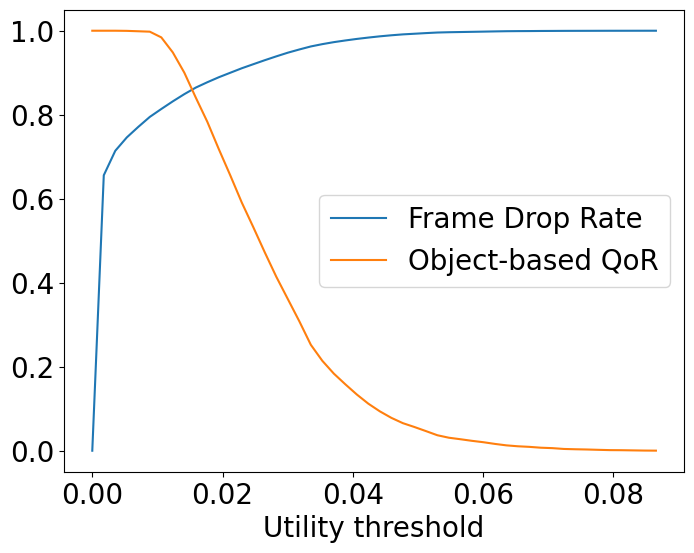}
        \caption{Variation of target object based QoR metric and frame drop rate with utility threshold for a \loadshedder{} tuned to detect red vehicles.}
        \label{fig:obj_cross_val_red}
    \end{subfigure}
    \caption{Performance of the utility-based \loadshedder{} on a query looking for Red cars as target objects.}
  \label{fig:red_only}
\end{figure*}

We begin by evaluating a query looking for target objects of a single color, which in this are red cars. \cref{fig:util_cross_val_red} shows the utility of positive and negative frames for the videos in our dataset. The results show that the utility function computes significantly higher utility value for positive frames than negative ones. This result is significant because it shows the performance of the utility function on unseen videos.
\par We demonstrate how such a separation in utility values between positive and negative frames is useful in being able to detecting target objects and maintaining a high QoR value while also shedding a significant fraction of (useless) frames. In \cref{fig:obj_cross_val_red}, we show the variation of frame drop rate and QoR metric with respect to utility threshold. As expected, an increasing utility threshold results in an increase in the frame drop rate, which also includes a small portion of useful frames containing target objects, and hence results in a drop in the QoR metric.

\begin{figure*}[t!]
    \centering
    \begin{subfigure}{0.31\textwidth}
        \includegraphics[width=\textwidth]{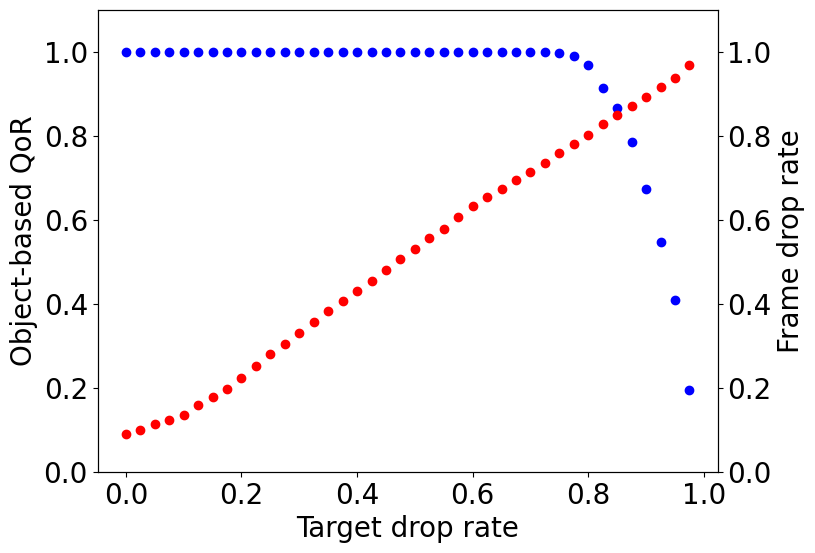}
        \caption{Impact of target drop rate on QoR rate and frame drop rate for utility-based approach.}
        \label{fig:util_drops}
    \end{subfigure}
    \hfill
    \begin{subfigure}{0.31\textwidth}
        \includegraphics[width=\textwidth]{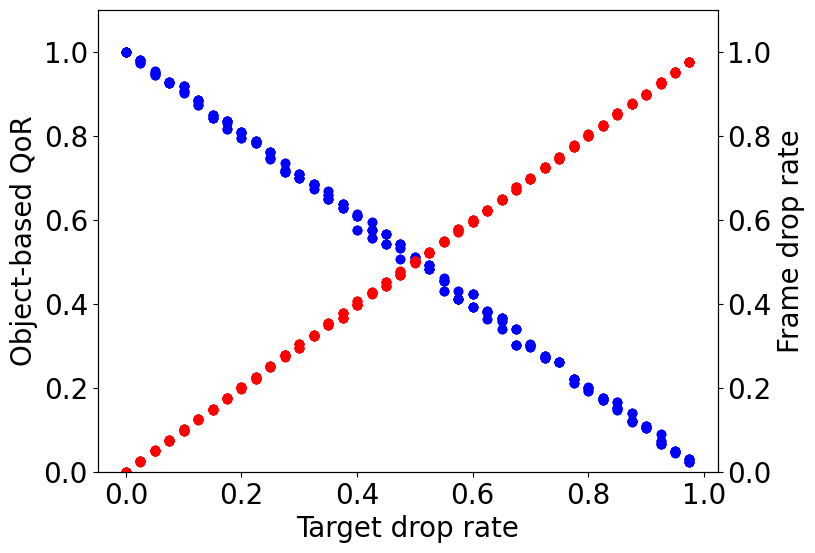}
        \caption{Impact of target drop rate on QoR and frame drop rate for content-agnostic approach.}
        \label{fig:random_drops}
    \end{subfigure}
    \hfill
    \begin{subfigure}{0.31\textwidth}
        \includegraphics[width=\textwidth]{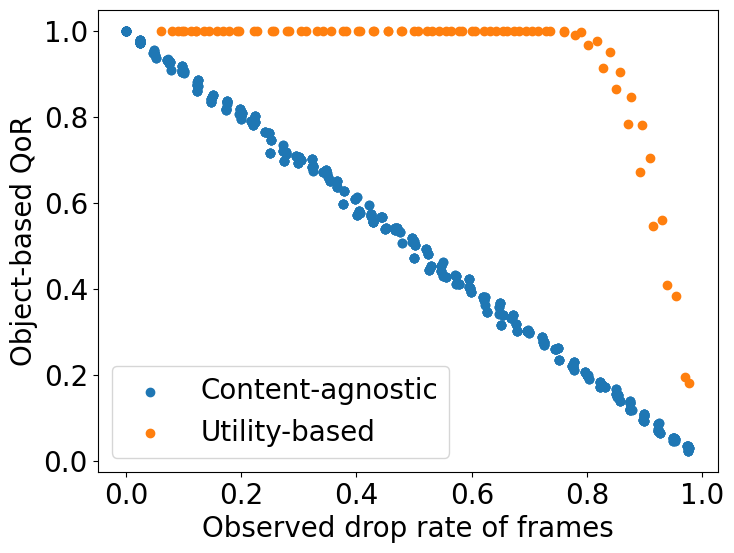}
        \caption{Comparison of tradeoff between QoR and frame drop rate for utility-based and content-agnostic load shedding approaches.}
        \label{fig:radom_red_tradeoff}
    \end{subfigure}
    \caption{Comparison of proposed utility-based load-shedding approach with a content-agnostic probability-based approach (random). \cref{fig:util_drops} and \cref{fig:random_drops} show the impact of target drop rate on the QoR metric and observed frame drop rate for utility-based and the content-agnostic approach respectively. Tradeoffs between the aforementioned two metrics are shown for both approaches in \cref{fig:radom_red_tradeoff}.}
    \label{fig:random_comparison}
\end{figure*}

\par \noindent \textbf{Comparison against Content-agnostic load shedding. } We now compare the performance of the proposed utility-based load shedding approach against a content-agnostic approach that sheds a fixed rate of incoming frames using a uniform probability. Firstly, \cref{fig:util_drops} shows the variation of frame drop rate and the associated fraction of objects detected against the target drop rate of the \loadshedder{}. As we have seen in \cref{fig:util_cross_val_red}, there is a significant portion of negative frames that have a low utility. Hence, a low target drop rate results in the selection of a utility threshold that drops many more frames than the target drop rate. However, since those frames are low-utility ones, the QoR remains at 1.0 until the target drop rate becomes so high that higher-utility frames (containing target objects) need to be dropped. That is when the QoR metric dips. Next, \cref{fig:random_drops} shows the variation of observed frame drop rate and QoR against target drop rate for the \textit{Content-agnostic} shedding approach. Since the shedding is based on a uniform random probability distribution, we repeat each setting 20 times. Even though the observed frame drop rate is roughly equal to the target drop rate, the QoR falls sharply because content-agnostic shedding often sheds frames containing target objects. Finally, \cref{fig:radom_red_tradeoff} compares the QoR that the proposed load shedding approach can achieve for a given observed frame drop rate with the  content-agnostic shedding approach. Although the QoR of the content-agnostic approach continuously falls with increasing drop rate, the QoR for utility-based approach has a visible drop only when the observed frame drop rate gets close to 1.0. The result shows that the utility-based approach is highly selective in picking frames to send to Backend Query Executor, and one can achieve a much higher QoR with it for a given observed frame drop rate compared to a content-agnostic shedding approach.

\subsubsection{Composite-color query: Red OR Yellow}

\begin{figure*}[t!]
    \centering
    \begin{subfigure}{0.6\textwidth}
        \includegraphics[width=\textwidth]{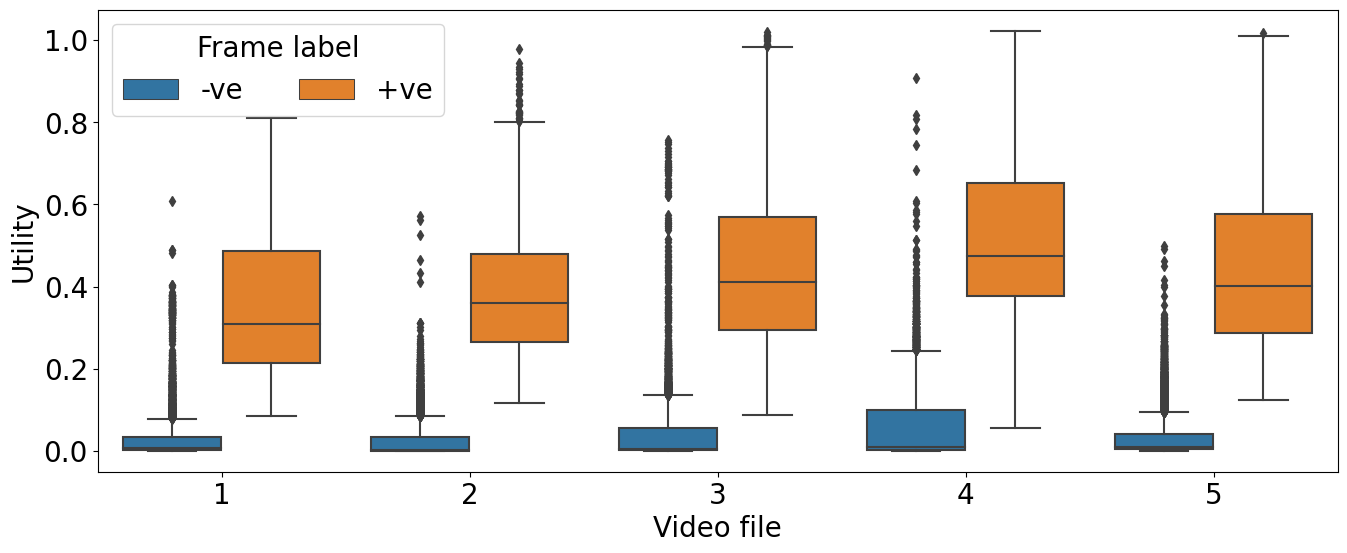}
        \caption{Utility values for positive and negative frames of unseen video frames. For a given video in the figure, the utility function used to compute its frames' utility values is trained using data that does not contain the given video.}
        \label{fig:util_cross_val_or}
    \end{subfigure}
    \hfill
    \begin{subfigure}{0.35\textwidth}
        \includegraphics[width=\textwidth]{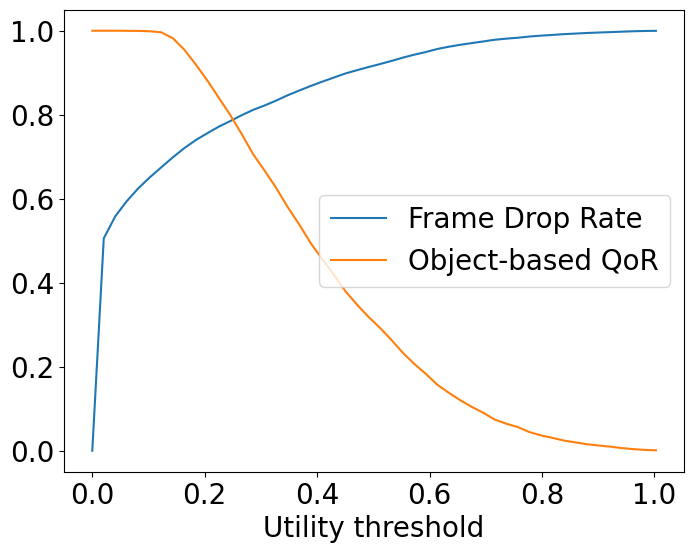}
        \caption{Variation of target object detection rate and frame drop rate with utility threshold for a \loadshedder{} tuned to detect red or yellow cars.}
        \label{fig:obj_cross_val_or}
    \end{subfigure}
    \caption{Performance of utility-based \loadshedder{} on query looking for Red OR Yellow cars as target objects.}
  \label{fig:or}
\end{figure*}

\begin{figure}[ht!]
  \centering  
  \includegraphics[width=0.95\columnwidth]{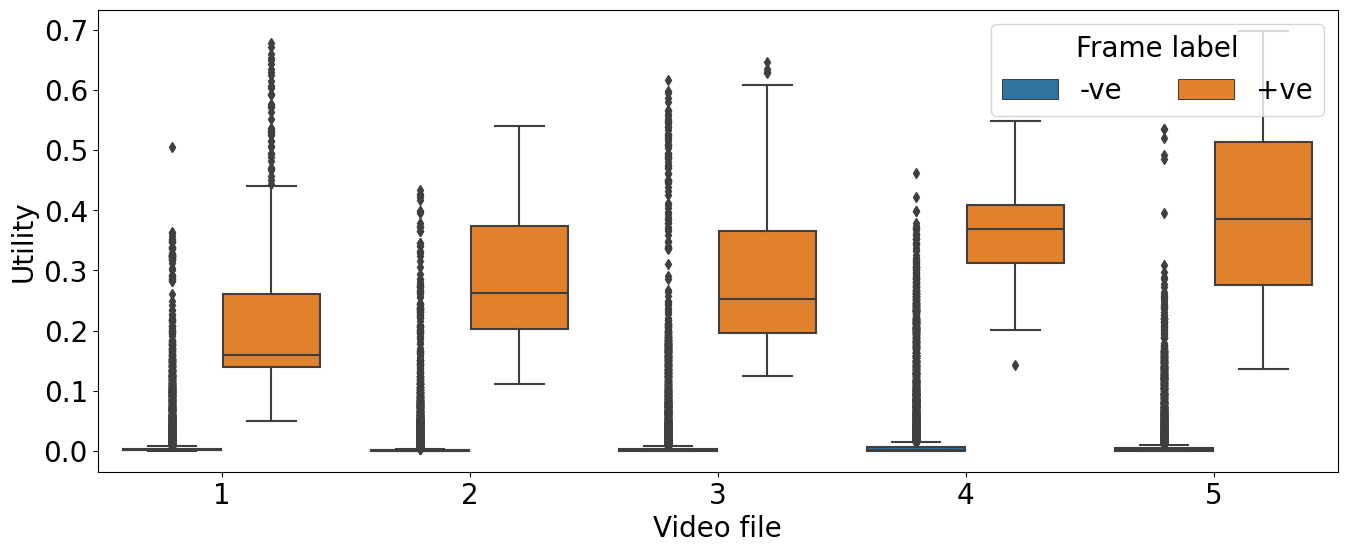}
  \caption{Utility values for positive and negative frames of unseen video frames. The given query is looking for frames containing both Red and Yellow cars.}
  \label{fig:util_cross_val_and}
\end{figure}

We perform a similar analysis for a composite queries - those that are tasked with (1) detecting target objects that are either Red or Yellow in color (we refer to this as the OR query), and (2) detecting all frames containing both Red and Yellow target objects (we refer to this as the AND query). As before, we iteratively select a set of videos as the training set and the complementary set as the test set. 
\par The utility value of frames for the OR query is shown in \cref{fig:util_cross_val_or}. Similar to the results for the single-color query, the utility value of positive frames is significantly higher than that of negative frames. Note that for the composite OR query, a positive frame is one that contains either a Red or Yellow target car. \cref{fig:obj_cross_val_or} shows the frame drop rate and QoR metric against the utility threshold. The QoR remains stagnant at 1.0 (selecting all frames containing target objects) with a high frame drop rate, until the utility threshold becomes high enough to start dropping positive frames. \cref{fig:util_cross_val_and} shows the utility value of frames for the AND query, and the differentiation between positive and negative frames is visible here as well. Note that for the composite AND query, a positive frame is one that contains both a Red or Yellow car.

\subsection{Application Evaluations}

\par The microbenchmarks evaluated the soundness of our utility calculation and compared it against a content-agnostic approach. In this subsection, we detail an E2E evaluation of running both the utility calculation and the \loadshedder{} control loop using a real-time video stream and show how it can timely control the latency of frame processing and avoid overloading the backend.

\subsubsection{Synthetic scenario}
\par First, we evaluate a synthetic worst-case scenario in which a sudden burst of high-processing activity occurs in the ingress video. The video comprises three segments: one low-utility frames with no target object, high-utility frames containing target object(s), and high-utility frames with no target object. To create such a tailored video, we obtain segments from the videos generated with Visual Road that are known a-priori to have those properties, and stitch them together to form a 15 minutes long video with each of the above three segments being 5 minutes long.
\par The expectation in this experiment is that during the video's first low-utility no-object segment, the \loadshedder{} will allow frames to be processed by the filter-stage in the backend query, even when the frame utility is low. This is because the the filter operator would drop these frames as they don't contain large blobs of the specific color that the query is looking for (because there are no target objects). Hence, the processing latency $proc_Q$ is low, leading to the throughput that can be handled by the backend higher than the incoming frame rate (\cref{eq:supported_tput}), and thereby a low target drop rate (\cref{eq:target_drop_ratio}). Next, in the second segment of the video, the \loadshedder{} will start shedding frames because all frames would be processed by the expensive DNN as they contain target objects. Thus, the \loadshedder{} increases the utility threshold so that the backend query executor can keep the end-to-end latency latency bounded for the frames that are processed. Finally, in the third segment with no target objects, the \loadshedder{} would stop shedding again and have an execution profile similar to the first segment. 

\begin{figure*}[ht!]
  \centering
  \begin{subfigure}{0.45\textwidth}
  \centering
  \includegraphics[width=0.95\columnwidth]{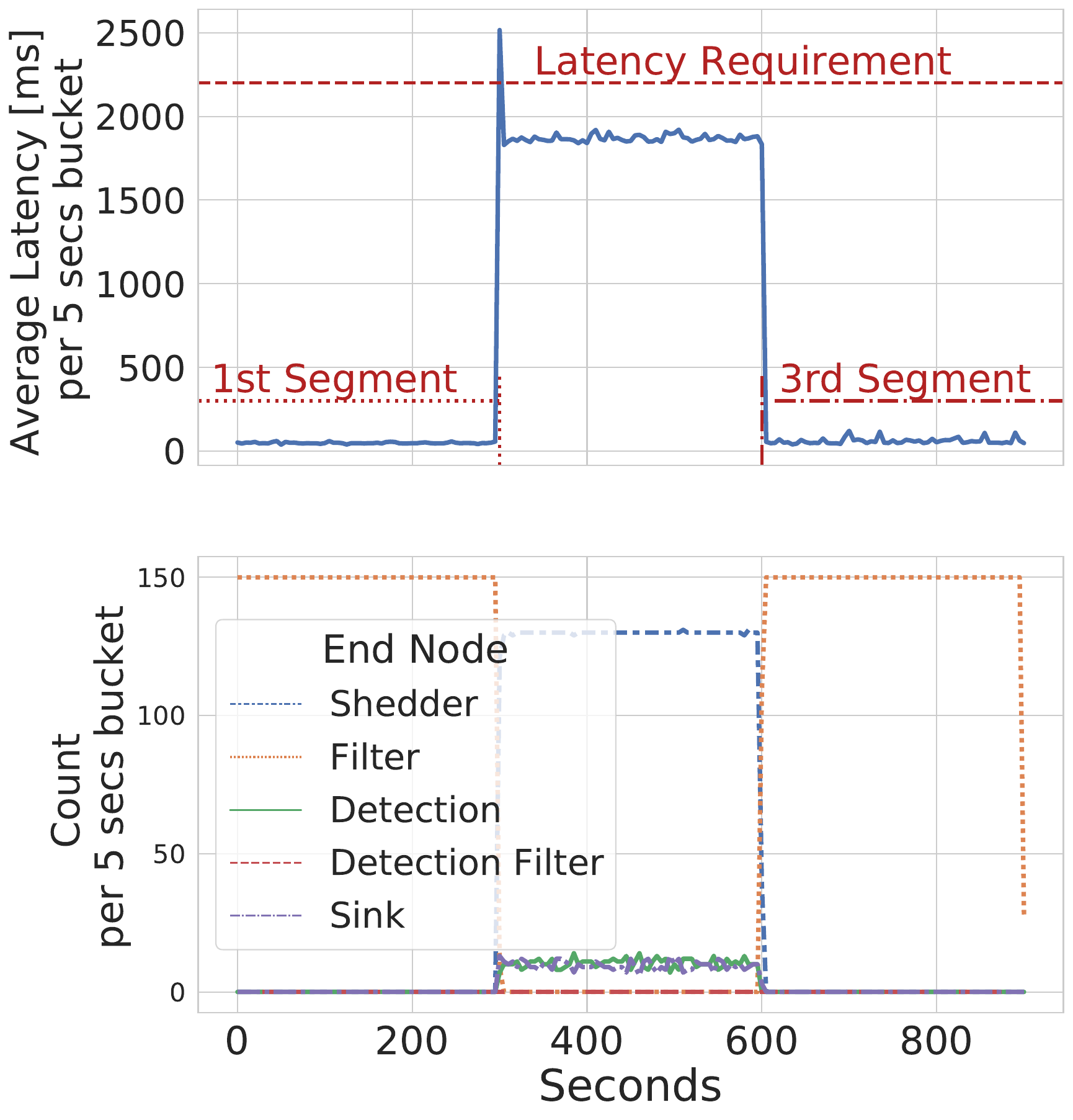}
  \caption{Synthetic Scenario each segments highlighted}\label{fig:synthetic_latency}
  \end{subfigure}
  \hfill
  \begin{subfigure}{0.48\textwidth}
  \centering
  \includegraphics[width=0.95\columnwidth]{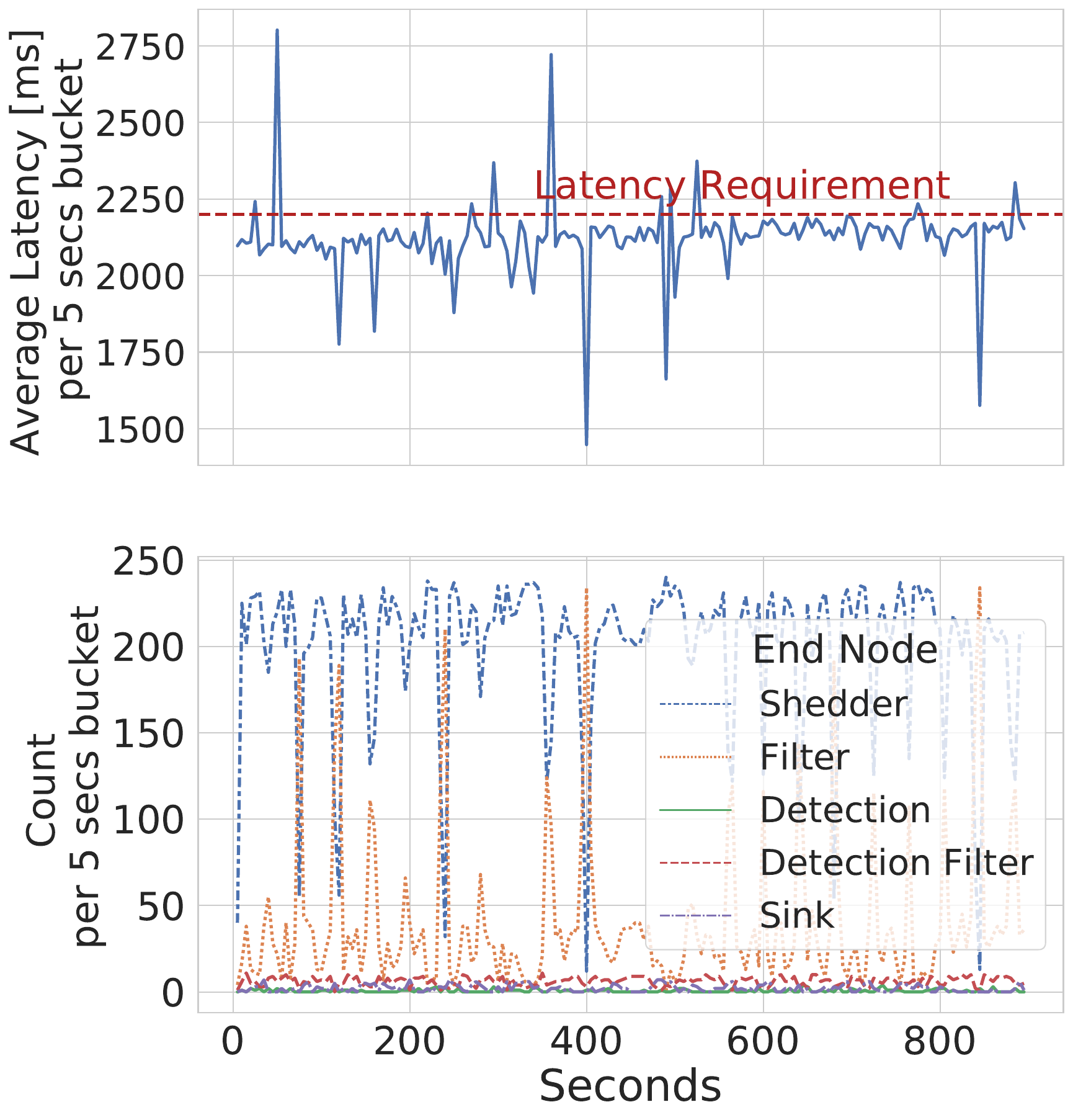}
  \caption{Realistic Scenario}\label{fig:realistic_latency}
  \end{subfigure}
  \caption{Analysis of both a synthetic and a realistic scenario. The upper graph of both figures shows in the y-axis the average processing latency grouped in five-second intervals, with the x-axis being the time in the video, highlighting the end-to-end latency requirement. The lower graph shows a breakdown of the number of frames that reached each query component (from \cref{fig:evaluation_diagram}); it shares the x-axis with the upper graph and uses the same five-second time intervals.}
  \end{figure*}

\par \cref{fig:synthetic_latency} shows the time-varying behavior of the query execution. The x-axis shows time, the y-axis in upper graph of \cref{fig:synthetic_latency} shows the max end-to-end latency in milliseconds for each 5 minutes time window, along with with the end-to-end latency requirement. The lower graph in \cref{fig:synthetic_latency} shows the number of frames processed at each stage group every 5 seconds. The upper graph in \cref{fig:synthetic_latency} shows how the latency is always less than the required end-to-end latency. Similarly, the lower graph in  \cref{fig:synthetic_latency} portrays how the \loadshedder{} decides the frame drop rate at each segment, with no shedding required in the first and third segment and plenty of shedding in the high-processing segment. Both these results match our expectations, and go on to show that the proposed \loadshedder{} can almost instantly react to changes and have a low number of violations even with extreme changes, that too for real-time video processing, where the input has usual a frame rate of 24 fps or less. There was only 1 latency violation during the peak in the second segment, while the \loadshedder{} was recalculating the queue size and the utility threshold.

\subsubsection{Realistic smart-city scenario}

\begin{figure}[ht!]
  \centering
  \includegraphics[width=0.9\columnwidth]{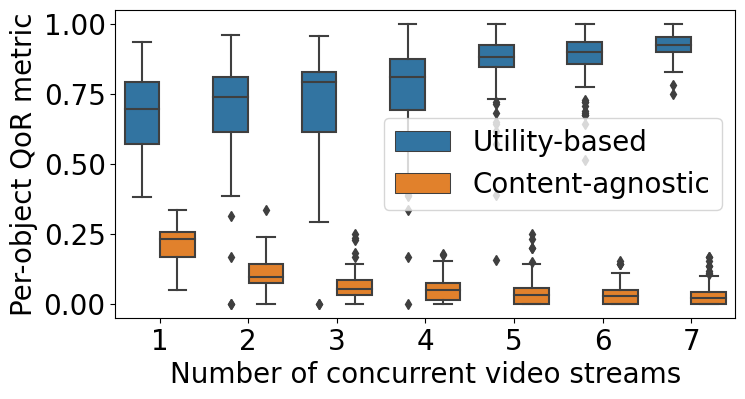}
  \caption{Comparison of per-object QoR metric of the proposed utility-based load shedding approach against a content-agnostic approach with fixed frame drop rate. The number of concurrent video streams fed into the \loadshedder{} is the control variable along the x-axis.}
  \label{fig:smart-city-accuracy}
\end{figure}

\par The previous sub-section analyzed the worst-case scenario for the \loadshedder{}. This sub-section analyzes the \loadshedder{} running directly on a subset of the videos generated using Visual Road. The Video Streamer generates a stream of frame features interleaved from multiple videos, thereby emulating the \loadshedder{} receiving frames from multiple cameras. We show the end-to-end latency and the distribution of frames processed at each stage, along with the variation  the QoR metric with an increasing number of concurrent video streams.  

\par Similarly to \cref{fig:synthetic_latency}, \cref{fig:realistic_latency} shows that the \loadshedder{} can keep the processing latency bounded by shedding some of the frames. There are more spikes than in the synthetic scenario because 
the DNN is invoked unpredictably by frames from different videos at different time periods. The \loadshedder{} is effective in minimizing the latency violations caused due to these sudden surges in load.
\cref{fig:realistic_latency} shows the latency and elements being processed for five concurrent videos. Additionally, \cref{fig:smart-city-accuracy} shows that the proposed utility-based load shedding approach can exploit the statistical multiplexing between multiple cameras' video streams and achieve a high QoR. On the other hand, a content-agnostic approach (that sheds frames with uniform probability) has poor QoR. We compute the target drop ratio for the latter approach using \cref{eq:supported_tput} and assuming that $proc_Q$ is 500 ms, which is a rather lenient assumption for the baseline.

\subsection{Runtime Overhead Analysis}
\label{sec:breakdown_analysis}
\begin{figure}[ht!]
  \centering
  \includegraphics[width=0.9\columnwidth]{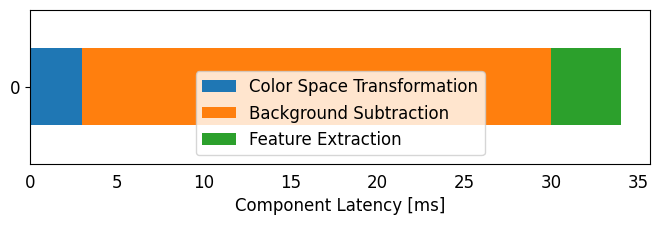}
  \caption{Breakdown of the additional latency incurred by the proposed load shedding approach. All reported values are median latencies. The utility calculation component is not shown here because its latency was negligible compared to the other components.}
  \label{fig:latency_breakdown}
\end{figure}

We evaluate the additional latency incurred by a video analytics system that includes the proposed load shedding approach as compared to executing the backend query without load shedding. We assume that cameras have colocated compute capability on them, which they use to do three tasks: (1) converting the color space of the source video stream from RGB to HSV, (2) subtracting the background from the frame to extract the foreground, and (3) extracting color features from the foreground which serves as the input to the \loadshedder{}. We evaluate the time taken to perform these tasks on a Nvidia Jetson TX1 with a quad-core ARM Cortex-A57 processor and 4 GB of RAM, which we use to represent the typical co-located compute power with cameras. We use a video stream with continuously high activity to stress test the above operations and obtain worst-case delay numbers. \cref{fig:latency_breakdown} shows the median latency incurred by each of the component tasks.
The overall additional latency remains below 35 ms which can support the video streams of multiple cameras operating at 10 frames per second (and even higher rates like the common 24 fps). With composite queries needing features calculated for multiple colors, the feature extraction portion of the overhead would be multiplied by the number of colors considered, while the other components' latencies would not change (as they are computed once per frame).
\section{Discussion}
\label{sec:discussion}
\par \noindent \textbf{Automatic selection of Hue ranges for a query.} The proposed load shedding approach requires minimal intervention of the application query developer, except having to provide the Hue range for the target objects. This task can also be automated by the analysis of bounding boxes of target objects available in the training data. Techniques such as dominant color detection \cite{dom_color} can be used to automatically extract the Hue ranges in target objects and fed into the utility calculation function.
\par \noindent \textbf{Feature calculation vs utility calculation on camera.} 
We could further improve the efficiency of the network by pushing the utility calculation itself to the camera. This decision will incur a tradeoff between the overhead of maintaining a distributed utility model versus a higher communication cost of sending all frames than those with a high utility value. When the cameras can calculate the utility, the load shedder can let them know what utility threshold to use to reduce unnecessary frames sent. However, when the model is updated due to either new data available or drift from the previous model, it needs to be updated at each camera, incurring additional bandwidth requirements. Therefore, this decision should be taken considering the connectivity scenario of the camera network.

\section{Conclusion}
\label{sec:conclusion}
In this paper we have shown how a low-cost \loadshedder can be built for real-time video processing on edge devices. The goal of the proposed \loadshedder{} is to shed frames such that the end-to-end processing latency of each video frame is under the latency bound for the query, while maximizing the quality of result (QoR). The \loadshedder{} uses color features of ingress video frames to compute a utility value, that denotes how likely a given frame is to contain a target object of the query at hand. We show how the proposed utility function is able to differentiate between video frames that do contain target objects from frames that don't. We propose a utility threshold based approach for the \loadshedder{} to enforce a target rate of dropping frames. The \loadshedder{} also consists of a control-loop component that continuously monitors the execution of frames and detects overload in the backend query. In the case of overload, it adjusts the target frame drop rate, which in turn sets a new utility threshold for frames to be compared against. 
\par Through our evaluations we have shown that the proposed \loadshedder{} is able to differentiate between frames containing target objects from those that don't. It is able to attain a high frame drop rate, wherein it drops as many "useless" frames as possible - thereby maintaining a high QoR metric. We show this result for queries looking for target objects of a single-color as well as composite queries looking for target objects of two possible colors. We show that the proposed \loadshedder{} is able to adjust the utility threshold dynamically according to the observed load on the backend query to meet the end-to-end latency constraint. Finally, we show that the overhead of executing the operations of the load shedding approach is not significant on edge devices.

\bibliographystyle{IEEEtran}
\bibliography{main_ieee}

\end{document}